\documentclass[oldversion,letterpaper]{aa}

\usepackage{txfonts}

\usepackage{graphicx, psfig, epsfig}

\usepackage{natbib}
\bibpunct{(}{)}{;}{a}{}{,} 

\usepackage{url}

\usepackage{aalongtable}
\setlongtables

\begin{document}
\title{Structure and flux variability in the VLBI jet of BL Lacertae during the
WEBT\thanks{For questions regarding the availability of WEBT data please contact
Massimo Villata (\email{villata@to.astro.it})} campaigns (1995--2004)}
\author{U.~Bach \inst{1}
\and M.~Villata \inst{1}
\and C.~M.~Raiteri \inst{1}
\and I.~Agudo \inst{2}
\and H.~D.~Aller \inst{3}
\and M.~F.~Aller \inst{3}
\and G.~Denn \inst{4}
\and J.~L.~G\'omez \inst{5}
\and S.~Jorstad \inst{6}
\and A.~Marscher \inst{6}
\and R.~L.~Mutel \inst{7}
\and H.~Ter\"asranta \inst{8}}
\offprints{U.\ Bach, \email{bach@to.astro.it}} 
\institute{Istituto Nazionale di Astrofisica, 
Osservatorio Astronomico di Torino, Via Osservatorio 20, 
10025 Pino Torinese (TO), Italy 
\and Max-Planck-Institut f\"ur Radioastronomie, Auf dem H\"ugel 69, 53121
Bonn, Germany
\and Department of Astronomy University of Michigan, 500 Church St. 830
Dennison, Ann Arbor, MI, 48109, USA
\and Metropolitan State College of Denver, Department of Physics, Campus Box
69, P.\ O. Box 173362, Denver, CO, 80217, USA
\and Instituto de Astrof\'isica de Andaluc\'ia (CSIC), Apartado 3004, 18080
Granada, Spain
\and Institute for Astrophysical Research, Boston
University, 725 Commonwealth Ave., Boston, MA 02215, USA
\and Department of Physics and Astronomy, 203 Van Allen Hall, University of
Iowa, Iowa City, IA 52242, USA
\and Mets\"ahovi Radio Observatory, Helsinki University of Technology,
Mets\"ahovintie 114, 02540 Kylm\"al\"a, Finland}
\date{Received 21 March 2006; accepted 29 May 2006}
\titlerunning{Structure and flux variability in the VLBI jet of BL Lacertae
during the WEBT campaigns (1995--2004)}
\authorrunning{U.\ Bach et al.}
\abstract{BL Lacertae has been the target of several observing campaigns by the Whole
Earth Blazar Telescope (WEBT) collaboration and is one of the best studied
blazars at all accessible wavelengths. A recent analysis of the optical and
radio variability indicates that part of the radio variability is correlated
with the optical light curve. Here we present an analysis of a huge VLBI data
set including 108 images at 15, 22, and 43\,GHz obtained between 1995 and 2004.
The aim of this study is to identify the different components contributing to the
single-dish radio light curves. We obtain separate radio light curves for the
VLBI core and jet and show that the radio spectral index of single-dish
observations can be used to trace the core variability. Cross-correlation of the
radio spectral index with the optical light curve indicates that the optical
variations lead the radio by about 100 days at 15\,GHz. By fitting the radio
time lags vs. frequency, we find that the power law is steeper than
expected for a freely expanding conical jet in equipartition with energy density
decreasing as the square of the distance down the jet as in the K\"onigl model.
The analysis of the historical data back to 1968 reveals that during a time
range of 16 years the optical variability was reduced and its correlation with the
radio emission was suppressed. There is a section of the compact radio jet where
the emission is weak such that flares propagating down the jet are bright first
in the core region with a secondary increase in flux about 1.0 mas from the
core. This illustrates the importance of direct imaging to the interpretation of
multi-wavelength light curves that can be affected by several distinct
components at any given time. We discuss how the complex behaviour of the light
curves and correlations can be understood within the framework of a precessing
helical jet model.}
\maketitle

\section{Introduction}

The term ``blazars'' identifies a family of radio-loud active galactic nuclei
(AGN) showing strong variability at all wavelengths from radio to $\gamma$-rays,
high degrees of polarization, and apparent brightness temperatures exceeding the
Compton limit (e.g., \citealt{1999APh....11..159U}). During the last decades a
rather general consensus on the global mechanism responsible for the emission
has been achieved: a supermassive black hole surrounded by a massive accretion
disk feeding a powerful jet closely aligned to the line of sight. Relativistic
electrons in the jet plasma produce the soft synchrotron photons (from radio
wavelengths to the UV band and sometimes even X-rays) through, while hard
photons (from X-rays to $\gamma$-rays) are likely produced by inverse Compton
scattering (e.g., \citealt{1996ApJ...463..444S} and references therein). Blazars
are divided into two subclasses: flat-spectrum radio quasars and BL Lac objects,
whose main difference is the lack or weakness of emission lines in BL Lac
objects (strongest rest-frame emission line width $<5$\,\AA; e.g.,
\citealt{1991ApJ...374..431S} and \citealt{1995PASP..107..803U} for a review).

BL Lacertae, located in the nucleus of a moderately bright elliptical galaxy at
a redshift of $z=0.069$ (\citealt{1978ApJ...219L..85M}) is the prototype of the
class of BL Lac objects. However, emission lines with equivalent width in excess
of $5$\,\AA\ have been detected several times
(\citealt{1995ApJ...452L...5V,1996MNRAS.281..737C,2000MNRAS.311..485C}),
suggesting that BL Lac objects can also have a broad line region (BLR), but it
would be usually outshone by the beamed synchrotron emission of the jet. BL Lac
has been studied intensively since its discovery and was the target of several
multi-wavelengths campaigns
(\citealt{1997ApJ...490L.145B,1999ApJ...515..140S,1999ApJ...521..145M,2002A&A...390..407V,2003ApJ...596..847B,2004A&A...424..497V,2004A&A...421..103V}).
In particular, the data collected during the Whole Earth Blazar Telescope
(WEBT)\footnote{http://www.to.astro.it/blazars/webt} campaigns provide an
unprecedented time sampling for radio and optical light curves from 1994 up to
now (\citealt{2002A&A...390..407V,2004A&A...424..497V,2004A&A...421..103V}).
Cross-corelation analysis of these light curves revealed well correlated
variability in the radio bands, where variations at higher frequencies lead the
lower-frequency ones by several days to a few months depending on the frequency
separation. The detection of a fair correlation between the optical $R$ band
variations and the ratio between the 22\,GHz and 5\,GHz radio flux densities
suggests that the optical emission and part of the radio emission arises from a
common origin in the inner portion of the jet. 

Very Long Baseline Interferometry (VLBI) observations of the parsec-scale
structure of BL Lac reveal a bright compact radio core and a jet extending to
the south (e.g., \citealt{1990ApJ...352...81M}). The jet emerges at a position
angle of P.A.~$\approx 195^\circ$ and displays a small bend at $\sim 4$\,mas
distance from the core towards P.A.~$\approx 160^\circ $. A number of
superluminal jet components have been observed displaying bent trajectories and
speeds from 3\,$c$ to $9\,c$ (e.g.,
\citealt{1990ApJ...352...81M,2000ApJS..129...61D,2003MNRAS.341..405S,2004ApJ...609..539K,2005AJ....130.1418J}).
Evidence for precession of the jet nozzle arising from periodic variations of
the electric vector position angle (EVPA) at millimetre wavelengths and the
position angle of newly emerging jet features (\citealt{2003MNRAS.341..405S})
are still under debate (\citealt{2005ApJ...623...79M}). A comparison of the
position angles of optical polarization vectors with the orientation of the
polarization vectors in 5\,GHz VLBI images suggests that the optical emission is
more likely to arise from the core than from the jet (\citealt{1994AJ....107..884G}).

Here we present an investigation of the flux density evolution of the parsec-
and subparsec-scale structure of BL Lac at 15\,GHz, 22\,GHz, and 43\,GHz using a
dataset of 108 VLBI observations performed between 1995 and 2004. The aim of
this study is to disentangle the different variability patterns that seem to be
present in the single-dish radio light curves and to identify the region of the
radio events that seem to correlate with the optical variability. Throughout
this paper we will assume a flat universe model with the following parameters:
Hubble constant $H_0=71$\,km\,s$^{-1}$\,Mpc$^{-1}$, a pressure-less matter
content $\Omega_{\rm m}=0.3$, and a cosmological constant $\Omega_{\rm
\lambda}=0.7$ (\citealt{2003ApJS..148..175S}). At the redshift of BL Lac this
corresponds to a luminosity distance of 308\,Mpc and to an angular resolution of
1.3\,pc/mas. Cosmology-dependent values quoted from other authors are scaled to
these parameters.

The observations and data reduction procedures will be described in
Sect.~\ref{observations}. Our results will be presented in Sect.~\ref{results}
and discussed in Sect.~\ref{discussion}. At the end we will give a summary with
the conclusions in Sect.~\ref{conclusions}.

\section{Observations and data reduction}\label{observations}

\subsection{Radio and optical light curves}

The optical and single-dish radio flux densities used in this work have already
been presented and analysed by \cite{2004A&A...424..497V,2004A&A...421..103V}.
The radio data at 4.8\,GHz, 8.0\,GHz, and 14.5\,GHz (labelled as 5, 8, and
15\,GHz throughout the article) were obtained by the AGN monitoring at the
University of Michigan Radio Astronomy Observatory
(UMRAO\footnote{\url{http://www.astro.lsa.umich.edu/obs/radiotel/umrao.html}};
\citealt{2003AAS...202.1801A}), while the 22\,GHz and 37\,GHz flux densities
were measured with the radio telescope of the Mets\"ahovi Radio Observatory
(\citealt{1998A&AS..132..305T,2004A&A...427..769T,2005A&A...440..409T}). The
optical $R$ band light curve was obtained by assembling data from various
observatories of the WEBT collaboration; details on the light curve construction
and the subtraction of the host galaxy contribution can be found in
\cite{2002A&A...390..407V,2004A&A...424..497V,2004A&A...421..103V}. Here we use
these data for a comparison with the structural evolution of the VLBI core and
jet structure on parsec and subparsec-scales.

\subsection{VLBI data}\label{sec:VLBIdata}

Most of the VLBI data used in this study were provided by the authors from
various large observing campaigns, where BL\,Lac was observed as a target
source\footnote{VLBA 2\,cm Survey
(\citealt{1998AJ....115.1295K,2002AJ....124..662Z,2004ApJ...609..539K}); MOJAVE
(\citealt{2005AJ....130.1389L}); \cite{2000ApJS..129...61D};
\cite{2001ApJ...556..738J,2005AJ....130.1418J}; \cite{2003MNRAS.341..405S};
\cite{2005ApJ...623...79M}} or as a polarization calibrator\footnote{Agudo et
al.\ \& G\'omez et al.\ priv.\ comm.\, target source published in
\cite{2000Sci...289.2317G,2001ApJ...561L.161G,Gomez2002}}. A list showing the
beam size, total flux density, peak flux density, rms noise level, and the
references where the data was first published for all our VLBI images is given
in Table~\ref{tab:obslog} (available in the electronic version). For the
details on the data reduction we refer to the  references of the observations.
The data were provided as calibrated ($u,v$)-data files and no further
self-calibration was applied. Three epochs (reference 6 in
Table~\ref{tab:obslog}) are newly reduced observations where BL Lac was used as
a polarization calibrator. Those were obtained with the full VLBA and the 100\,m
Effelsberg telescope at 15\,GHz and were correlated at the VLBA correlator in
Socorro, NM. The post-correlation analysis was done using NRAO's Astronomical
Image Processing System (AIPS). After loading the data into AIPS, the standard
amplitude and phase calibration was performed. Both imaging and phase- and
amplitude self calibration was done in Difmap (\citealt{1997adass...6...77S}),
using the CLEAN (\citealt{1974A&AS...15..417H}) and SELFCAL procedures. All data
were imaged with the same image parameters at each frequency using uniform
weighting in Difmap. Images with matching cell-sizes and resolution (high
frequency data were tapered to the lower frequency resolution) were produced for
the spectral analysis. The analysis of the images, including spectral index
maps, slices and flux density measurements, was done in AIPS.

\begin{figure*}[htbp]
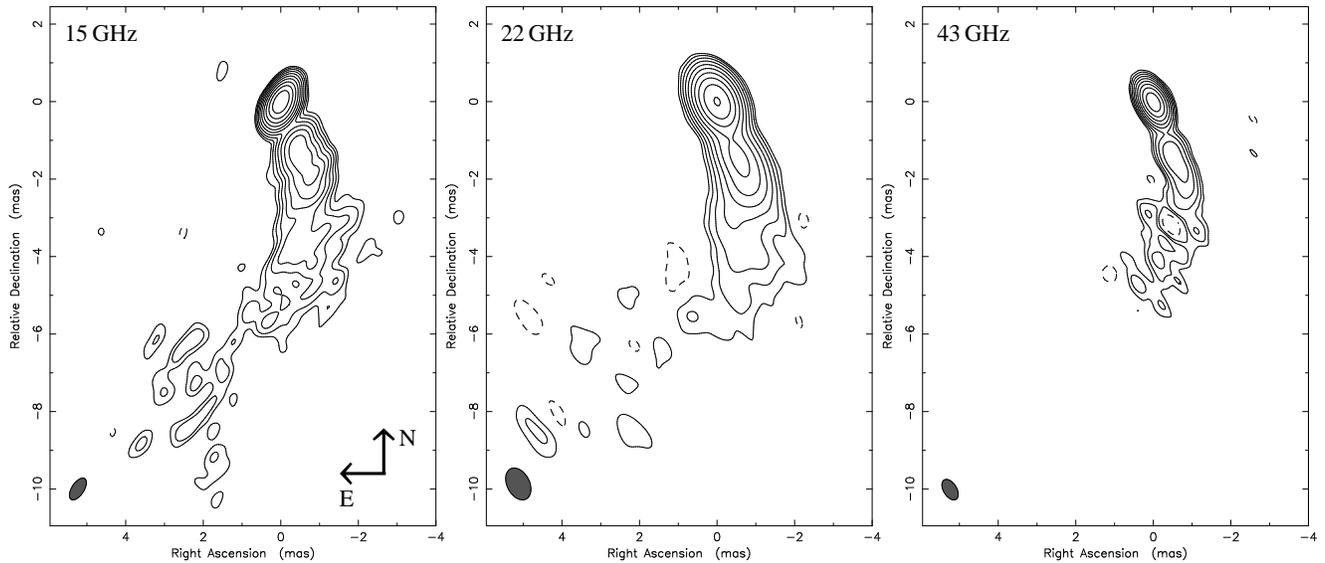

\vbox{\centering
\includegraphics[bb= 67 100 557 732,angle=0,width=5.8cm,clip]{5235_f01a.ps}
\put(-130,200){\makebox(0,0){15\,GHz}}
\put(-23,42){\makebox(0,0){\huge$\uparrow$}}
\put(-14,47){\makebox(0,0){N}}
\put(-31,33){\makebox(0,0){\huge$\leftarrow$}}
\put(-37,24){\makebox(0,0){E}}
\includegraphics[bb= 67 100 557 732,angle=0,width=5.8cm,clip]{5235_f01b.ps}
\put(-130,200){\makebox(0,0){22\,GHz}}
\includegraphics[bb= 67 100 557 732,angle=0,width=5.8cm,clip]{5235_f01c.ps}
\put(-130,200){\makebox(0,0){43\,GHz}}}
\caption{VLBI images of BL~Lac. {\bf Left:} 15\,GHz
image
(epoch 2002.03) with a beam of 0.64\,mas $\times$ 0.32\,mas at P.A. $-33^\circ$. Peak flux
density is 1.4\,Jy/beam and contours start at 1.5\,mJy/beam, increasing by a
factor of 2. {\bf Middle:} 22\,GHz image
(2002.05) with a beam of 0.88\,mas $\times$ 0.58\,mas at P.A. $27^\circ$. Peak flux
density is 1.6\,Jy/beam and contours start at 3\,mJy/beam, increasing by a
factor of 2. {\bf Right:} 43\,GHz image
(2002.05) with a beam of 0.58\,mas $\times$ 0.34\,mas at P.A. $32^\circ$. Peak flux
density is 1.3\,Jy/beam and contours start at 3\,mJy/beam, increasing by a
factor of 2.}\label{fig:BLimages}
\end{figure*}

Sample images of BL Lac at 15\,GHz, 22\,GHz, and 43\,GHz are shown in
Fig.~\ref{fig:BLimages}. The jet is visible up to 10\,mas (13\,pc) from the
bright VLBI core and bends at about 4\,mas from a position angle of
$\sim195^\circ$ to $\sim160^\circ$ (measured counterclockwise from north). The VLBI
core is clearly distinguished from the jet by a region of low emission (between 0.7\,mas
and 1.0\,mas core distance) most of the time. This becomes more evident in
Fig.~\ref{fig:slices}, where the source intensity profiles along
P.A.~$=195^\circ$ are shown (see next section for more details). 

\section{Results}\label{results}


\subsection{VLBI light curves}

\begin{figure}[htbp]
\centering
\includegraphics[bb= 80 20 570 752,angle=-90,width=7.5cm,clip]{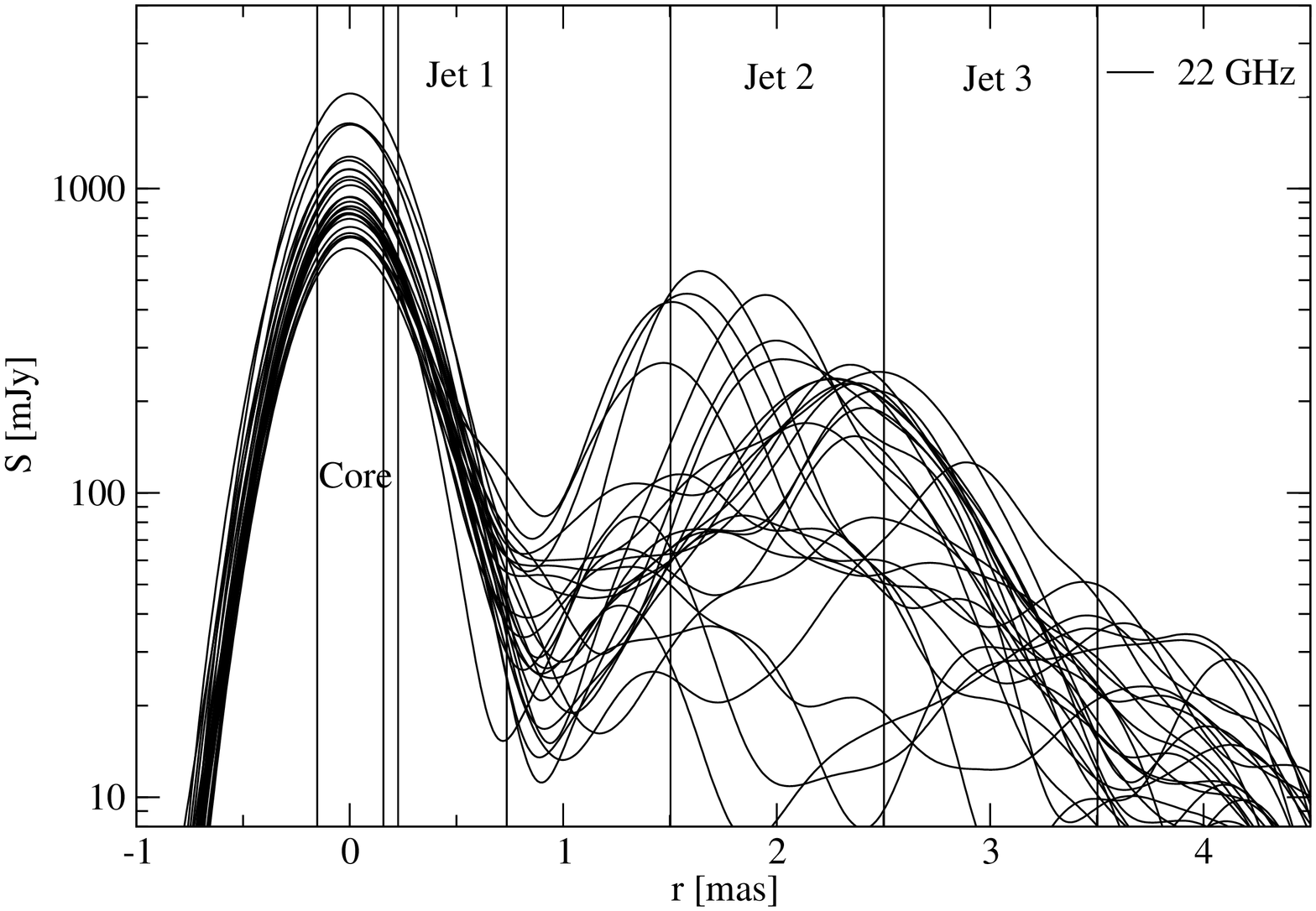}
\includegraphics[bb= 70 20 570 752,angle=-90,width=7.5cm,clip]{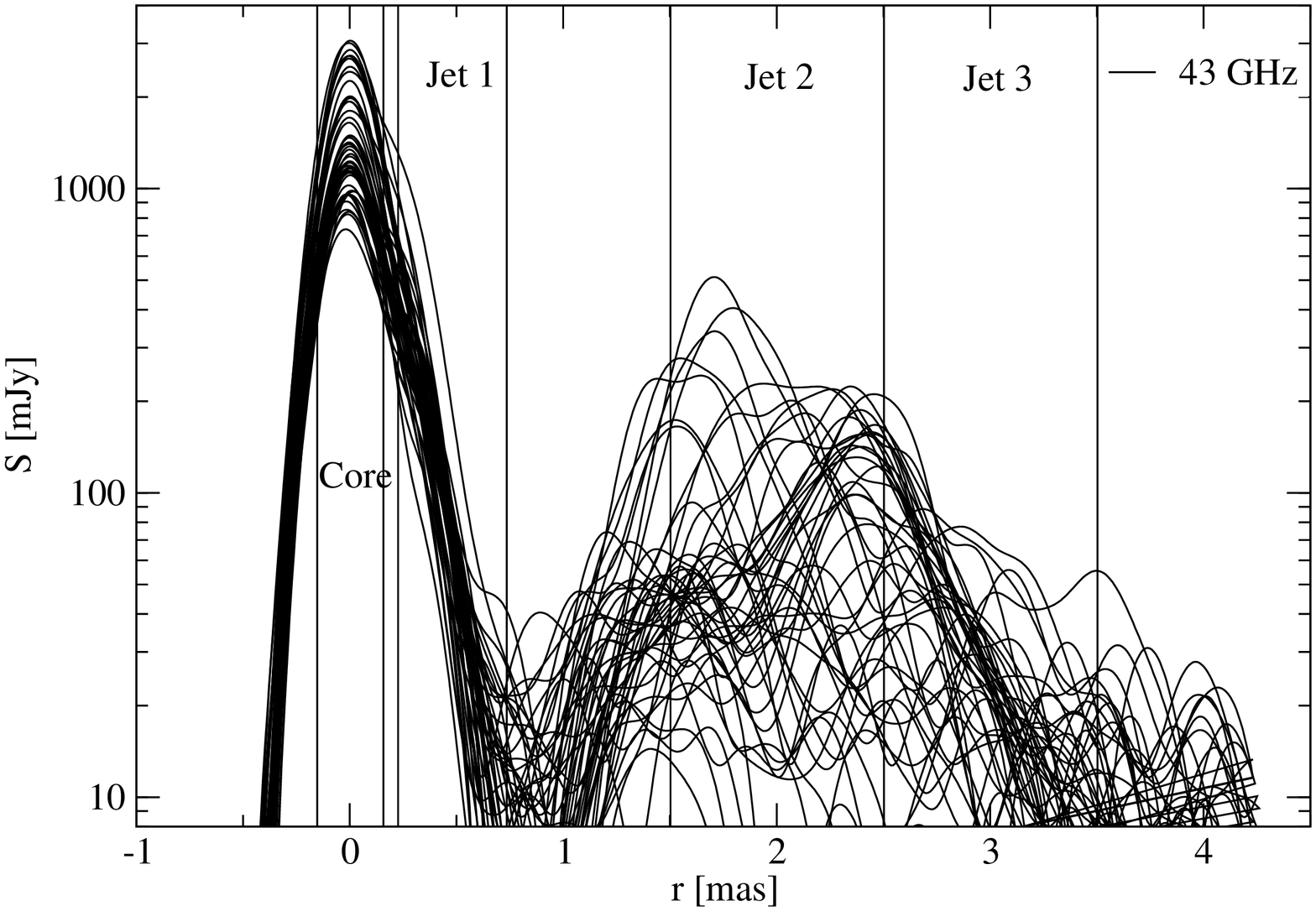}
\caption{Intensity profiles of the VLBI images of BL\,Lac at 22\,GHz (top) and
43\,GHz (bottom), combining all epochs. The figure illustrates the flux density
behaviour of the jet ridge-line  at a resolution of 0.5\,mas (22\,GHz) and
0.3\,mas (43\,GHz) at P.A.~$=195^\circ$. Each curve represents one epoch
between 1997 and 2003. Vertical lines indicate the core and jet regions that
were used to analyse the images.}\label{fig:slices}
\end{figure}

To quantify the variability of different regions of the source we extract
light curves of the core and of different portions of the jet. The regions that
we chose are displayed in Fig.~\ref{fig:slices}. Based on the jet kinematics we
expect that a flare which occurs in the core will propagate along the jet and
cause the jet to brighten at a certain position as the feature passes by. We
will later see that this is indeed the case. Therefore, the exact location of
the flux density measurement along the jet seems  not very important, since a
flare that travels down the jet will at some time pass by. Remarkable in this
context is the low emission region around 0.7--1\,mas from the core, where only
very small variations occur even though the core and the jet around 2\,mas are
highly variable and several new jet components were reported during this time
range
(\citealt{2000ApJS..129...61D,2001ApJ...556..738J,2003MNRAS.341..405S,2004ApJ...609..539K,2005AJ....130.1418J}).
It is obvious that a measurement in this region would give completely different
results than a measurement 0.5\,mas farther down the jet. From
Fig.~\ref{fig:slices} we can estimate the maximum brightness that can be reached
at a certain jet position. It seems that new components first rapidly fade as
they separate from the core and then reappear at about 1\,mas distance from the core.
The maximum brightness is reached at about 1.5\,mas and after that the
components slowly fade as they travel down the jet.

\begin{figure*}[htbp]
\centering
\includegraphics[bb= 40 40 550
765,angle=-90,width=18cm,clip]{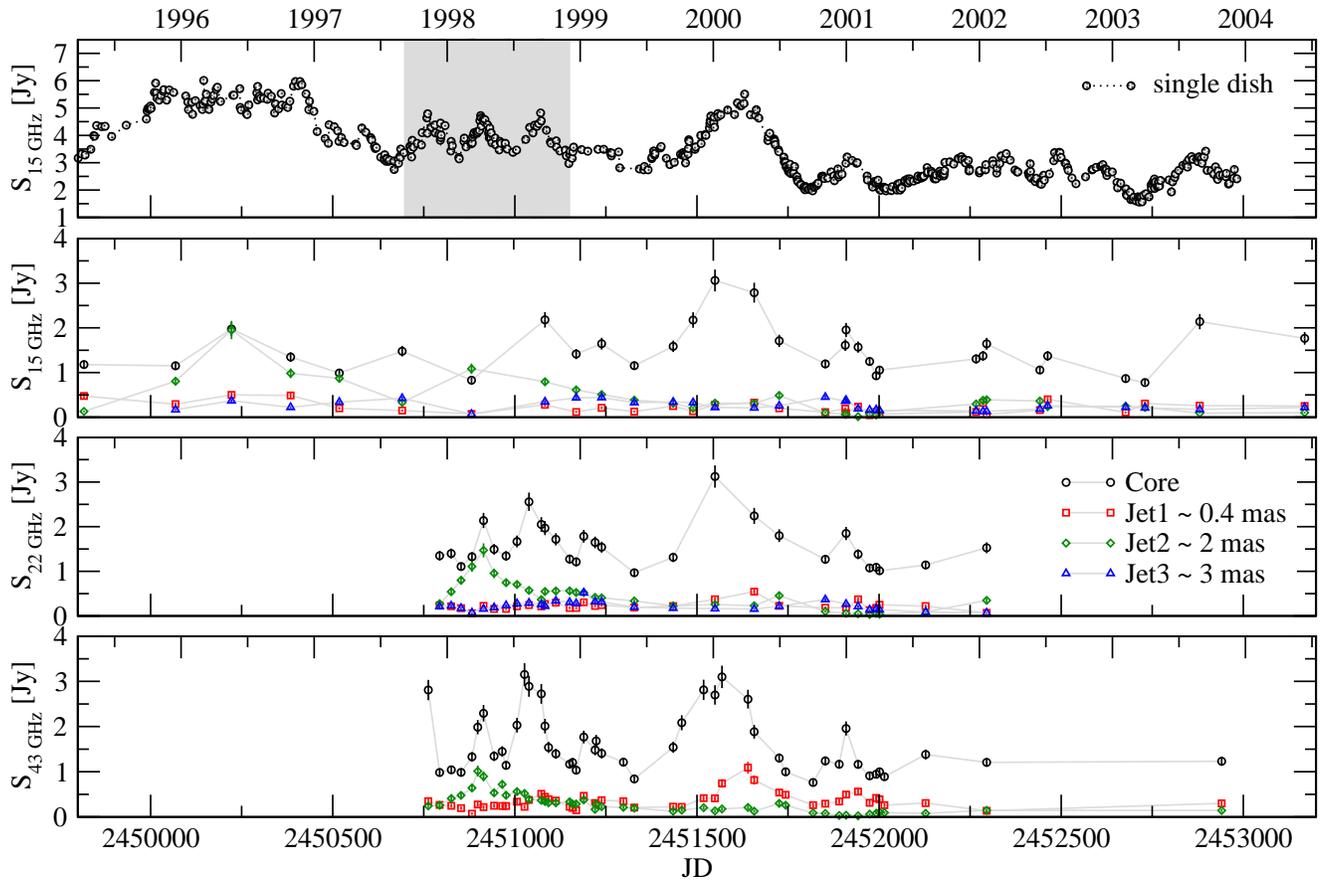}
\caption{Flux density evolution of different portions of the VLBI core-jet
structure of BL\,Lac at 15\,GHz,
22\,GHz, and 43\,GHz compared with the 15\,GHz single-dish light curve.
Different symbols denote different separation from the core $r$ (see text for
more details). The grey box highlights an example of a nice case study, where
there are three equally bright flares, but the VLBI data reveals that contrary
to the others the middle one was dominated by the jet and not by the
core. Note that due to the rapid fading of the jet we do not show a
Jet3 curve at 43\,GHz.}\label{fig:VLBAvstime}  
\end{figure*}

We have tested several alternative positions with different sizes to check how the
light curves change at different core distances. If we take measurements
within a very small region, the integrated flux density goes down and the signal to
noise ratio of the light curve decreases. On the other hand, if the region is too
large, different jet components will overlap and smear out the variability. We
find that the best choice is a small region close to the core, where
flares from the core might be detectable with only a short time delay, as well as
two slightly larger regions after the low emission gap to trace the flux density
evolution farther downstream.

The core is represented by a circular region of 0.15\,mas radius around the
image centre, the first jet region corresponds to a core separation of
$0.2\,{\rm mas}\leq r \leq 0.7$\,mas, the second one to $1.5\,{\rm mas}\leq r
\leq 2.5$\,mas, and the third one to $2.5\,{\rm mas}\leq r \leq 3.5$\,mas. We
have tested several methods to obtain the corresponding light curves, including
1) the task IMSTAT in AIPS, which integrates the flux density of the image
within a predefined rectangular region, 2) model fitting of Gaussian components
in Difmap; and 3) integrating the flux density along the slices presented in
Fig~\ref{fig:slices}. Among the alternatives, we find that a quite accurate and
flexible method is to extract the light curves by summing the delta-function
components within the specified regions from the final CLEAN models produced
with Difmap. Another advantage of this method is that one can easily change the
limits of the regions and rerun the analysis. Since most of the data were
reduced by various authors and the quality of the data is inhomogeneous, we
estimate the flux density error to be 10\%, which is typical for VLBI
measurements. From our experience with the VLBA, 10\% is a conservative
assumption and the typical uncertainty at 15\,GHz and 22\,GHz might be as low as
5\%. Our three VLBA+Effelsberg epochs (2002.03, 2002.51, and 2003.24) at 15\,GHz
have typical errors of 5\% estimated from the gain corrections during the
amplitude self-calibration.

Figure~\ref{fig:VLBAvstime} shows the final VLBI light curves in comparison with
the 15\,GHz single-dish light curve from UMRAO. One can clearly see that the
single-dish light curve is usually dominated by variability of the VLBI core,
but occasionally parts of the jet reach or even overcome the core brightness.
The latter is most evident at 15\,GHz before mid 1998.
\cite{2004A&A...424..497V} calculated the flux density ratio (a ``hardness
ratio,'' which is comparable with the radio spectral index) between the 22\,GHz
and 5\,GHz light curves and distinguished two different kind of flares. Flares
that are more pronounced at higher frequencies are called hard flares (inverted
or flat radio spectrum), while flares that are dominant at lower frequencies are
called ``soft flares'' (steep radio spectrum).

From a comparison of Fig.~\ref{fig:VLBAvstime} with Fig.~8 in
\cite{2004A&A...424..497V}, one can see that the hard flares are observed when
the VLBI core is bright compared to the jet, while during soft phases the jet is
relatively bright. For example, the second flare (1998.2) of the three equally
bright flares in the single-dish light curve between 1997.7 and 1998.9 in
Fig.~\ref{fig:VLBAvstime} (grey box) is softer than the adjacent ones (cf.\ also
Fig.~8 in \citealt{2004A&A...424..497V}). If we take a look at the VLBI light
curves it becomes obvious that, although in the single-dish measurements the
flares have a similar brightness, the 1998.2 flare has a large contribution from
a brightening in the jet at about 2\,mas core distance (jet2) and originates not
only from the core. It seems therefore very important to know the jet
contribution to the single-dish light curves if we intend to compare the radio
variations with X-ray or optical light curves, since they are most likely
dominated by the emission from the jet-base and not from the outer optically
thin jet emission.

\subsection{Radio spectral index}

\begin{figure}[htbp]
\centering
\includegraphics[bb= 20 20 777 1177,angle=0,width=9cm,clip]{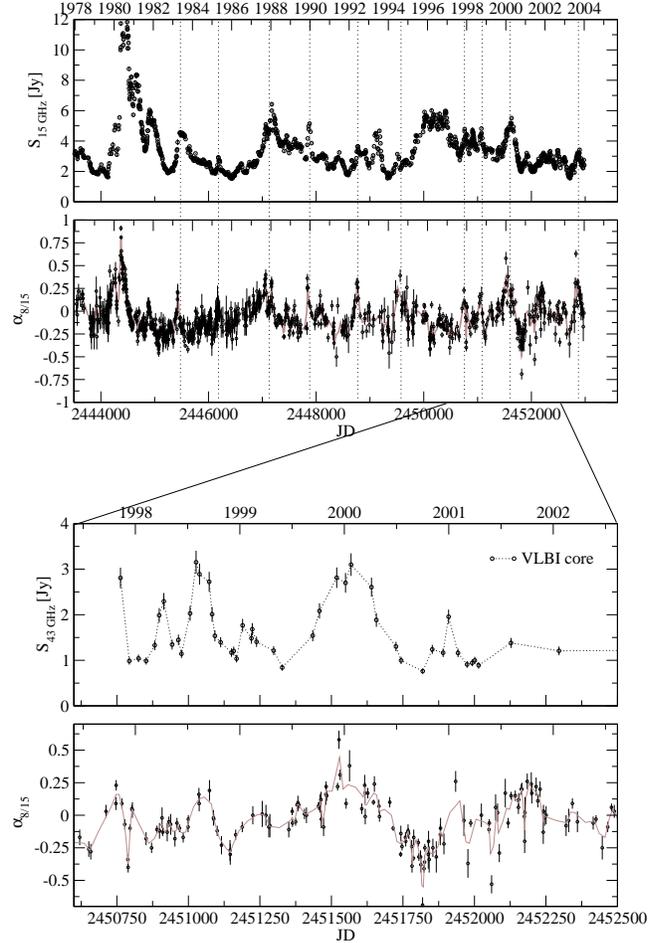}
\caption{{\bf Top panels:} Radio light curve at 15\,GHz in comparison with the
spectral-index curve between 8\,GHz and 15\,GHz. One can see that most of the
flares are accompanied by a flattening of the spectrum. The vertical dotted
lines mark some examples and help to guide the eye. {\bf Bottom panels:}
Zoom into the time range of the VLBI monitoring at 43\,GHz. The grey line
represents a three point average of the spectral-index curve. There is a
close apparent correlation between the 43\,GHz core flux density and 
$\alpha_{8/15}$ (see also Fig.~\ref{fig:sp-core_dcf}).}\label{fig:spix}
\end{figure}

The single-dish radio spectral index $\alpha$ (defined as $S_\nu \propto
\nu^\alpha$, where $S_\nu$ is the flux density at frequency $\nu$) was
calculated for all adjacent frequency pairs between 5\,GHz and 37\,GHz with
separation shorter than 3~days. This limit is usually fulfilled for data from
the same observatory (5, 8, 15\,GHz from UMRAO and 22 and 37\,GHz from
Mets\"ahovi) and therefore reduces the sampling significantly only for the
15\,GHz to 22\,GHz spectral index. The mean time sampling for the radio light
curves is typically of 10~days, but after 1990 the time sampling becomes much
better ($\sim 5$~days). The spectral indices obtained for different pairs of
frequencies (5/8, 8/15, 15/22, and 22/37 GHz) all follow the same behaviour in
time. In Fig.~\ref{fig:spix} (top panel) we show the spectral-index curve
between 8\,GHz and 15\,GHz ($\alpha_{8/15}$), which is the best sampled one. One
can see that most of the flares are accompanied by a flattening of the spectrum.
The flattening seems to peak shortly before the flare reaches its maximum. A
comparison between the 43\,GHz VLBI core flux density and $\alpha_{8/15}$ is
shown in the bottom panels. The single-dish spectral index is able to trace the
VLBI core variability in a very accurate way. We will test this using a discrete
correlation function (DCF) in the next section.

\begin{figure}[htbp]
\centering
\includegraphics[bb= 20 20 565 842,angle=-90,width=9cm,clip]{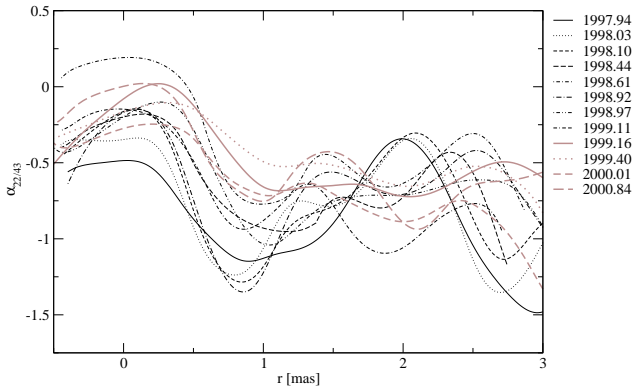}
\caption{Ridge-line profiles of the spectral index maps between 22\,GHz and
43\,GHz showing the core and the inner 3\,mas of the jet along P.A.~$195^\circ$.
The original images were tapered to the same resolution and convolved with a
circular beam of 0.7\,mas.}\label{fig:spix_slice} 
\end{figure}

The time evolution of the spectral index of the VLBI core and the inner 3\,mas
of the jet can be seen in Fig.~\ref{fig:spix_slice}. Shown are the ridge-line
profiles of BL Lac across the spectral index VLBI maps between 22\,GHz and
43\,GHz along P.A.~$195^\circ$. The spectral index maps were produced by
aligning the two images on the brightest component. To reduce the effect of a
potential core shift between the two frequencies, we convolved the images with a
0.7\,mas circular beam. Since there are no strong spectral gradients visible in
the core region that would indicate a core shift, we assume that the effect is
negligible at the resolution of our images. The core spectrum in
Fig.~\ref{fig:spix_slice} is significantly flatter than the jet for most of the
time especially when it is flaring (e.g., epoch 1998.61). It therefore seems
reasonable that the variability of the single-dish radio spectral index is a good
tracer for the variations in the VLBI core.

\subsection{Cross-correlation}\label{sec:cross}

A cross-correlation of the single-dish light curves of BL Lac has already been
performed by \cite{2004A&A...424..497V} using a DCF analysis
(\citealt{1988ApJ...333..646E,1992ApJ...386..473H}). We will continue to use
this method here. The authors found that the higher frequency radio variations
lead those at lower frequencies by several days to a few months, depending on
the frequency separation. In particular, the variations at the lowest frequency
(5\,GHz) lag the other frequencies by $\sim 40$ days (8\,GHz), $\sim 59$ days
(15\,GHz), and $\sim 78$ days (22\,GHz) with uncertainties of 5 to 10 days.
These values were derived by calculating the centroid of the DCF near its peak.
The corresponding uncertainties were estimated from variations of the time
binning and the calculation of the centroid for cutoffs between 70\% and 80\% of
the peak value. To complete the frequency coverage, we derive here also the DCF
for the 5\,GHz and 37\,GHz frequency pair and find a time lag of
$105\pm15$ days. To be consistent with \cite{2004A&A...424..497V} all DCFs in
this work are performed for the same time range, between JD=2446500 (March
1986) and JD=2453000 (December 2003), which excludes the highly pronounced
radio flare in 1980. Such prominent events could lead to spurious peaks in the
DCF analysis, since a peak would appear at every time lag at which it overlaps
with a flare in the second light curve. We note that the time range that we use
is shorter for the VLBI data, which are only available between 1995 and 2004.
All DCF calculations were performed several times with different binning in time
and over different time ranges. The plots shown here correspond to the versions
that showed the most robust results. In most cases small changes did not affect
the results significantly and, where larger changes were observed, these are
discussed in the text.

\begin{figure}[htbp]
\centering
\includegraphics[bb= 30 20 590 762,angle=-90,width=9cm,clip]{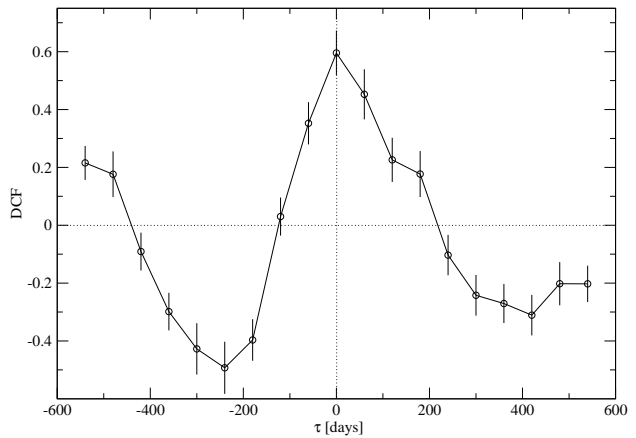}
\caption{DCF of the 43\,GHz VLBI core flux density (1997--2002) and the
radio spectral index between 8\,GHz and 15\,GHz showing a good correlation with a
possible short delay of the spectral index.}\label{fig:sp-core_dcf}
\end{figure}

\begin{figure}[htbp]
\centering
\includegraphics[bb= 40 20 590 762,angle=-90,width=9cm,clip]{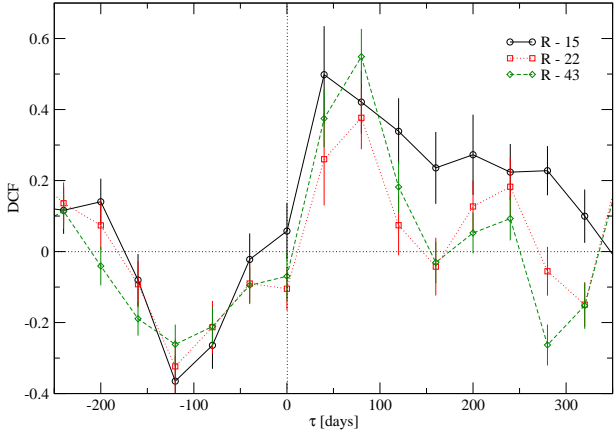}
\caption{DCF of the optical  $R$ band light curve with the VLBI core light curves at
15\,GHz (1995--2003), 22\,GHz and 43\,GHz (1997--2002).}\label{fig:RDCF}
\end{figure}

\cite{2004A&A...424..497V} found a modest correlation between the optical light
curve and the high-frequency radio ones with a radio time lag of about 100 days.
Moreover they showed that a better correlation is found when comparing the
optical variability with the spectral variations in the radio bands, which
highlights radio flares that are more pronounced at higher frequencies. On the
other hand, the radio spectral index and the VLBI core flux density
(Fig.~\ref{fig:spix}) also seem to agree very well. The corresponding DCF
analysis supports the correlation (Fig.~\ref{fig:sp-core_dcf}). There appears
also a possible anti-correlation at about $-250$ days, but this is very likely
due to the broad peak around 2000.0 overlapping with the dip at $sim 2000.7$
(Fig.~\ref{fig:spix}). Calculation of the centroid of the highest 3 points of
the DCF gives a short time delay of about 4 days for the spectral index, but at
a bin size of 60 days the uncertainties are too large to reliably measure such
short delays. We thus performed cross-correlations between the optical $R$ band
light curve\footnote{We use optical flux densities corrected for the host galaxy
contribution (\citealt{2002A&A...390..407V,2004A&A...424..497V})} and our VLBI
core light curves. The results are shown in Fig.~\ref{fig:RDCF}. The time
sampling of the VLBI data is not as good as for the single-dish light curves.
This might explain why the correlation is not so strong; however, all curves
show a tentative correlation with a 50 to 100 day time lag. The strongest
correlation is found with the 43\,GHz core light curve, which is also the best
sampled one.


\begin{figure}[htbp]
\centering
\includegraphics[bb= 40 20 572 762,angle=-90,width=9cm,clip]{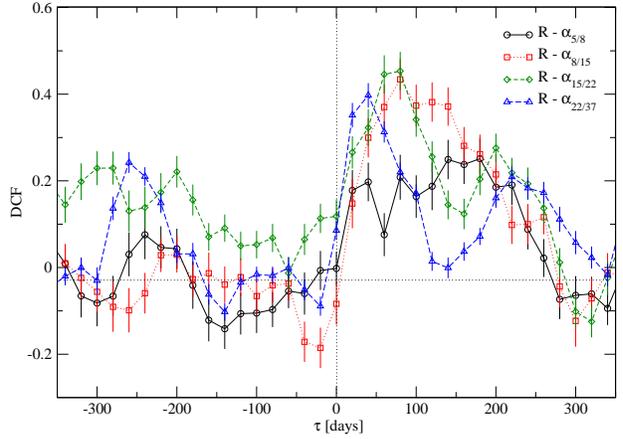}
\includegraphics[bb= 60 20 592 762,angle=-90,width=9cm,clip]{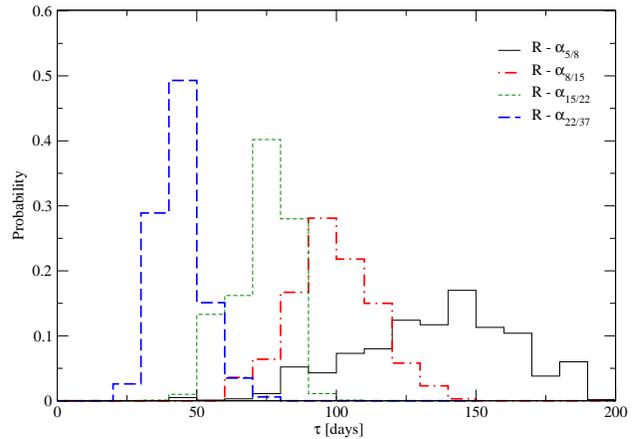}
\caption{{\bf Top panel:} DCF of the optical $R$ band light curve with all
spectral index pairs between 5\,GHz and 37\,GHz (1986--2003). Although the
correlations reach a maximum peak value of only about 0.5, there is a clear
trend to shorter time delays for higher frequencies. {\bf Bottom panel:}
Probability plot for the DCF peaks resulting from the correlation between
optical data and radio spectral indices derived from 1000 Monte Carlo
simulations.}\label{fig:RSPDCF}
\end{figure}

\begin{figure}[htbp]
\centering
\includegraphics[bb= 30 20 778 1136,angle=0,width=9cm,clip]{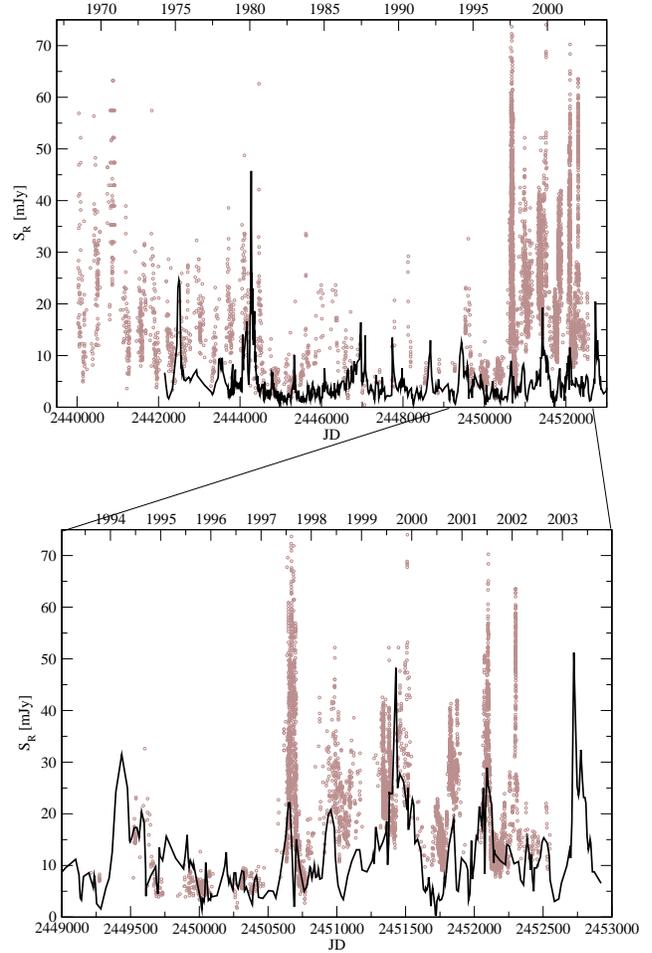}
\caption{Optical light curve vs. radio spectral index ($\alpha_{8/15}$). The
spectral index is shifted by --99 days. To make the comparison easier we
represent the spectral index by $(\alpha_{8/15}+0.8)^3$ (top) and
$(\alpha_{8/15}+0.8)^4$ (bottom), so that all values are positive and the peaks
are enhanced. The curve is smoothed with a three point average to reduce the
noise. {\bf Top:} All data. {\bf Bottom:} Only the last 10 
years with the best sampled data.}\label{fig:Rspix}
\end{figure}

We obtain a better result from the cross-correlation of the optical data with
the spectral indices. Figure~\ref{fig:RSPDCF} (top panel) shows the DCF for all
four frequency pairs between 5\,GHz and 37\,GHz (1986--2003). Although the DCF
peaks are below 0.6, the clear trend for shorter time delays at higher frequency
pairs is consistent with the radio cross-correlations, which show that the
higher frequencies lead the lower ones (\citealt{2004A&A...424..497V}). The
secondary peaks with negative time delays most likely come from the regular
pattern of optical flares after 1997, visible in Fig.~\ref{fig:Rspix}, which
appear at separations of a bit less than 1\,year (6 flares in 5 years).
Auto-correlations of both the optical light curve and the 22/37\,GHz spectral
index weakly show this periodicity if we analyse only the time-range after 1997,
but the periodicity vanishes outside of this 5-year period. Therefore, we cannot
rule out that the light curves of BL\,Lac sometimes show periodicities, but then
it is only a short-term phenomenon. Periodicity analyses of BL Lac light curves
performed by various authors have found periods between 2.5\,yr and 4\,yr, about
8\,yr, and 15\,yr to 20\,yr for the radio (e.g.,
\citealt{2003ApJ...591..695K,2004A&A...424..497V}), and about 1\,yr, 5\,yr, 7\,
to 8\,yr, 10\,yr, and 15\,yr in the optical (e.g.,
\citealt{1997ARep...41..154H,2004A&A...424..497V}), at different significance
levels. It is evident that further investigations are needed before we can draw
any conclusions regarding the presence of true, persistent periodicities.

The formal statistical significance of the found time lags using the DCF
analysis in this work is not easy to asses. In general the statistical
significance of a correlation depends on the number of points that contribute to
the calculation of the time lag (e.g, \citealt{2001sac..conf....3P}). Since the
number of points that contribute to the peak value of the DCF in our analysis
always exceeds 100, all peaks higher than 0.25 have to be considered as being
significant at a confidence level of about 98\%. Therefore we carefully checked
by eye how many flares with which amplitude in the light curves contribute to
the correlation and if there are any dominant flares which could strongly affect
the DCF (see also Fig.~\ref{fig:Rspix}). Hence we might call the 
$R-\alpha_{22/37}$ DCF with a peak value of 0.4 better than the correlation
between $\alpha_{8/15}$ and the 43\,GHz VLBI core light curve with a peak value
0.6, because the larger time interval and better sampling of $R-\alpha_{22/37}$
correlation makes it more reliable. A summary of the used parameters for all
DCFs is given in Table~\ref{tab:DCF}.

Another method to test the robustness of the DCF peaks is to perform Monte Carlo
simulations in which the curves are randomly modified within the measured
uncertainties and only subsets of the data are correlated -- a technique known
as ``flux redistribution/random subset selection'' (FR/RSS;
\citealt{1998PASP..110..660P}; see also \citealt{2003A&A...402..151R}). We used
this method to test the peaks of the $R--\alpha$ cross-correlation
(Fig.~\ref{fig:RSPDCF} (top)) through 1000 Monte Carlo simulations. In
Fig.~\ref{fig:RSPDCF} (bottom) the fraction of occurrence of the 75\% centroid
values of all Monte Carlo realizations for the four DCFs are given. The
histogram shows that the time delay and the uncertainty of the DCF between the
$R$ band and the radio spectral index decreases towards higher radio
frequencies.

\setcounter{table}{1}
\begin{table}[htbp]
 \caption{Summary of cross-correlation parameters. Columns contain the
frequency pairs (numbers denote radio frequencies in GHz and a ``V'' marks data obtained
from VLBI images), time interval of correlation (MJD=JD-2400000), binning
interval, DCF peak value, time-lag of the peak, time lag of the 75\% centroid, and a reference to the
corresponding figure.}
\centering
\scriptsize
 \label{tab:DCF}
  \begin{tabular}{lcccrrc} \hline
   \multicolumn{1}{c}{type} &
   \multicolumn{1}{c}{time int.} &
   \multicolumn{1}{c}{bin} &
   \multicolumn{1}{c}{DCF} &
   \multicolumn{1}{c}{$\tau_{\rm peak}$} &
   \multicolumn{1}{c}{$\tau_{\rm cent.}$} &
   \multicolumn{1}{c}{Fig.} \\
    &
   \multicolumn{1}{c}{[MJD]} &
   \multicolumn{1}{c}{[days]} &
    &
   \multicolumn{1}{c}{[days]} &
   \multicolumn{1}{c}{[days]} &
    \\ \hline 
$\alpha_{8/15}$--43V   & 50760--52950 & 60 &  0.60 &  0 &  0 & 6\\
$R$--15V               & 49800--53100 & 40 &  0.50 & 40 & 58 & 7\\
$R$--22V               & 50760--51080 & 40 &  0.38 & 80 & 64 & 7\\
$R$--43V               & 50760--52600 & 40 &  0.55 & 80 & 61 & 7\\
$R$--$\alpha_{5/8}$    & 46500--52600 & 20 &  0.26 &180 &148 & 8\\
$R$--$\alpha_{8/15}$   & 46500--52600 & 20 &  0.41 & 80 & 93 & 8\\
$R$--$\alpha_{15/22}$  & 46500--52600 & 20 &  0.45 & 80 & 71 & 8\\
$R$--$\alpha_{22/37}$  & 46500--52600 & 20 &  0.39 & 40 & 39 & 8\\
  \end{tabular}
\end{table}

To summarize: a fair correlation was found between $S_{\rm core}$ and the
radio spectral index with no measurable time delay, which suggests that the
radio spectral index is a good indicator for the VLBI core variability. A
tentative direct correlation is seen between $S_{\rm core}$ and the optical
light curve with radio time lags of less than 100 days. This, together with the
correlation between the optical flux and the radio spectral index, provides
evidence that the optical variability is indeed correlated with the radio
emission of the VLBI core.

In Table~\ref{tab:Rbandlags} we summarize the estimated time delays for the five
radio frequencies during the time range of 1986 and 2003. Since the flattening
of the spectrum appears when the higher frequency light curve rises earlier
and/or faster than the lower frequency one, the variability of the spectral
index should represent the core variability at the higher frequency. Therefore,
we attribute the time lag between the optical emission and the spectral index to
the highest radio frequency. The delay of the 5\,GHz variations, the lowest
frequency light curve in the analysis, is derived by adding the time lag between
the 5\,GHz and 8\,GHz radio light curves to the $R$--8\,GHz time lag. This
procedure might lead to small shifts of the derived time lags, since the shape
and peak of the spectral index variations are a combination of both frequencies,
but tests with different frequency pairs that are more widely separated (e.g.,
15/37\,GHz or 8/37\,GHz) support this approach and yield shifts of less than 5
days. This is well within the 1\,$\sigma$ errors derived from the Monte Carlo
realizations.

\begin{table}[htbp]
 \caption{Time lags, $\tau$, measured between the optical and the radio
variations. The values are derived from the correlation of the optical emission
with the radio spectral indices (see text for details).}
\centering
 \label{tab:Rbandlags}
  \begin{tabular}{lc} \hline
   \multicolumn{1}{c}{opt -- $\nu$ [GHz]} &
   \multicolumn{1}{c}{$\tau$ [days]} \\ \hline 
$R$ -- 5  & $175\pm30$ \\
$R$ -- 8  & $136\pm28$ \\
$R$ -- 15 & $ 99\pm15$ \\
$R$ -- 22 & $ 71\pm10$ \\
$R$ -- 37 & $ 44\pm10$ \\ \hline
  \end{tabular}
\end{table}

Figure~\ref{fig:Rspix} illustrates the good agreement between the optical light
curve and the radio spectral index ($\alpha_{8/15}$). In this figure the
spectral index is shifted by --99 days according to the $R$--15 delay
(Table~\ref{tab:Rbandlags}). Moreover, to make the comparison easier, we plot
the spectral index by $(\alpha_{8/15}+0.8)^3$ (top) and $(\alpha_{8/15}+0.8)^4$
(bottom), so that all values are positive
and the peaks are enhanced. One can see that the curves agree fairly well during
the last $\sim10$ years (bottom panel).

\section{Discussion}\label{discussion}

\subsection{Optical-radio time delays}\label{sec:timetest}

The long observational history of BL\,Lac led to several attempts to search for
radio-optical correlations. Some of these found correlations with radio time
delays of several hundred days
(\citealt{1976AJ.....81..489P,1992ApJ...386..473H,1994A&A...289..673T,1995AJ....110..529C,2002A&A...394...17H}).
Sometimes simultaneous events were reported
(\citealt{Balonek1982,1994A&A...286...80T}). A common result is that there are
periods where radio and optical variations correlate better and others where the
correlation seems to disappear. The correlations found here
(Sect.~\ref{sec:cross}) are calculated for data from 1986 to 2003, thus
excluding the large outburst in 1980. Although the sampling is not homogeneous,
it seems that the optical emission underwent a period of suppressed activity
from 1981 to early 1997 (Fig.~\ref{fig:Rspix}), during which flares were less
frequent and not so pronounced. Such a period is visible neither in the radio
light curves nor in the radio spectral index (Fig.~\ref{fig:spix}).

By constraining the DCF analysis on smaller time ranges we test how the time
delays change with time. This was done exemplarily for the cross-correlation of
the R band fluxes with  $\alpha_{8/15}$. The data between 1986 and 1989 are not
so well sampled, but although the DCF results become less significant, they
confirm the previous findings. During the period before 1986 a broad peak at
about 300 days is found, which is in good agreement with the results from the
studies of data up to 1992
(\citealt{1976AJ.....81..489P,1992ApJ...386..473H,1994A&A...289..673T,1995AJ....110..529C,2002A&A...394...17H}).
During the period of weak optical variability (from 1986 to early 1997) a small
peak around zero time delay indicates some simultaneous events
(\citealt{Balonek1982,1994A&A...286...80T}), but in general the DCF during this
time shows no peak above 0.25. Finally the most recent data (1997 to 2003),
characterized by the high optical activity, are dominated by a peak at around
100 days. The fact that both \cite{2004A&A...424..497V} and our new analysis
detected this 100-day time delay for data obtained from 1986 until now,
demonstrates that the most recent correlation seems to dominate over that found
in the previous period of weak variability. However, the time range used here is
not dominated by a major outburst, but rather consists of several flares in the
optical (6--7 major events after 1997 and about the same number of minor flares
before 1997), and the variability of the spectral index displays nearly the same
amplitude throughout the whole period. Therefore, the optical-radio delays found
in Sect.~\ref{sec:cross} seem to represent the result of a true correlation in
the variability behaviour of BL\,Lac over at least the last 10 years of our
dataset.

\subsection{Frequency dependence of the delay}

Frequency dependent time delays in radio flares are commonly related to a
combination of optical depth and travel time along the jet. Evolving flares
first appear in the inner jet region at the highest frequencies and, as the
perturbations (VLBI components) travel down the jet, the flares become visible
at successively lower frequencies where the optical depth of synchrotron
self-absorption decreases to $\tau_{\rm s} \sim 1$ at that frequency. For a
given magnetic field strength ($B(r)$) and electron energy distribution scale
factor ($N(r)$), the corresponding $\tau_{\rm s}$ is (e.g.,
\citealt{1998A&A...330...79L}):

\begin{equation}
\tau_{\rm s}=C_2(\alpha)N_1\left(\frac{eB_1}{2\pi m_{\rm e}}\right)^\epsilon
\frac{\delta^\epsilon \phi_0}{r^{(em+n-1)}\nu^{\epsilon+1}}
\end{equation}

Here $m_{\rm e}$ is the electron mass, $e$ is the electron charge,
$\epsilon=3/2-\alpha$, $\alpha$ is the optically thin synchrotron spectral
index, and $m$ and $n$ are the power-law exponents of the radial distance
dependence of the magnetic field and $N(r)$, respectively (e.g.,
\citealt{1998A&A...330...79L}). The observed jet opening angle is $\phi_0=\phi
\csc \theta$. The factor $C_2(\alpha)$ is described in
\cite{1970RvMP...42..237B}, and $C_2(-0.5)=8.4\cdot 10^{10}$ in cgs units.
Setting $\tau_{\rm s}$ to unity gives us an estimate of the distance from the
convergence point of the jet to the core as observed at frequency $\nu$.

Hence, we should be able to measure a frequency-dependent core position in VLBI
images: for a conical jet geometry the shift is given by $r \propto
\nu^{-1/k_{\rm r}}$, where $k_{\rm r}=[(3-2\alpha)m+2n-2]/(5-2\alpha)$ (e.g.,
\citealt{1998A&A...330...79L}). For the case of equipartition between  the
magnetic field and electron energy densities, the simplest choice of $m=1$ and
$n=2$ (\citealt{1981ApJ...243..700K}) leads to $k_{\rm r}=1$ independent of the
spectral index (\citealt{1998A&A...330...79L}). Larger values are reached
in regions with steeper pressure gradients than if, e.g., the electrons suffer
adiabatic energy losses without reacceleration (\citealt{1980ApJ...235..386M}).
Thus, taking measurements of the core shift  between several frequencies allows
to estimate $k_{\rm r}$ along the jet (e.g.,
\citealt{1998A&A...330...79L,2004A&A...426..481K}).

If the radio time lags are also due to opacity effects and the flare travels at
a constant speed, then they should also be proportional to $\nu^{-1/k_{\rm r}}$.
Recent kinematic studies on high frequency VLBA data of BL\,Lac have consistently
derived jet speeds with a bulk Lorentz factor of about 7 and an angle to the
line of sight of  $\sim12^\circ$, with only minor deviations
(\citealt{2000ApJS..129...61D,2003NewAR..47..641R,2003MNRAS.341..405S,2004ApJ...609..539K,2005AJ....130.1418J}).
We therefore assume that the Lorentz factor and angle to the line of sight are
more or less constant, and that the time delays are proportional to the distance
between the two emitting regions. E.g., a time delay,  $t'$, of 10 days would
then correspond to a distance, $r$, of about 0.06\,pc ($r=t\,\beta\,c$,
with $\beta=\sqrt{1-\frac{1}{\gamma^2}}$ and $t=\gamma\,t'$).

\begin{figure}[htbp]
\centering
\includegraphics[bb=  30 40 575 822,angle=-90,width=9cm,clip]{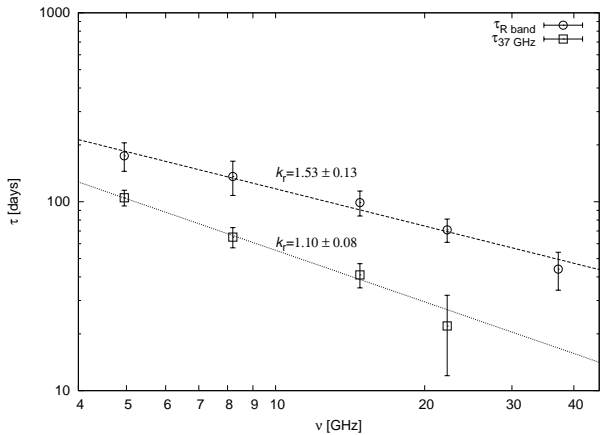}
\caption{Measured time lags {vs.} frequency. The circles and
dashed line represent the measured time lags and the corresponding fit with
respect to the optical emission, while the squares and dotted line denote
time lags and fit with respect to the 37\,GHz light curve.}\label{fig:time lags}
\end{figure}

In Fig.~\ref{fig:time lags} we plot the measured radio lags with respect to both
the 37\,GHz data and to the optical data {vs.} frequency. A power-law fit
represents the data reasonably well in both cases (reduced $\chi^2 < 1$) and
there it seems that the assumption of a constant jet speed holds at least in the
radio regime. It is important to note that the value of $k_{\rm r}$ strongly
depends on the absolute time lag, i.e. on the distance to the base of the jet,
which might be not far from the central engine. Although the optical emission
could be emitted close to the foot point of the jet, we should consider that
this region is still at some considerable distance from the central engine. In
this case $k_{\rm r}$ would be larger. If, on the other hand, the jet speed
between the radio and the optical emitting regions is not constant, the simple
assumption of $\tau \propto \nu^{-1/k_{\rm r}}$ is invalid. However, it seems
that at least in the radio regime the speed is constant and we can infer that
$k_{\rm r}$ is likely larger than 1.1. This implies that the gradients in jet
parameters are steeper than in the case of equipartition with $m=1$. Our result
is, on the other hand, consistent with a freely expanding jet with
adiabatic energy losses, in which case $n=(2/3)(3-2\alpha)$
(\citealt{1980ApJ...235..386M}) and $k_{\rm r}=(15-14\alpha)/[3(5-2\alpha)]$ for
$m=1$, or $k_{\rm r}\approx 1.2$ for $\alpha \approx -0.5$.

\subsection{Connection with jet components}

\begin{figure*}[htbp]
\centering
\includegraphics[bb= 30 20 572 772,angle=-90,width=15.5cm,clip]{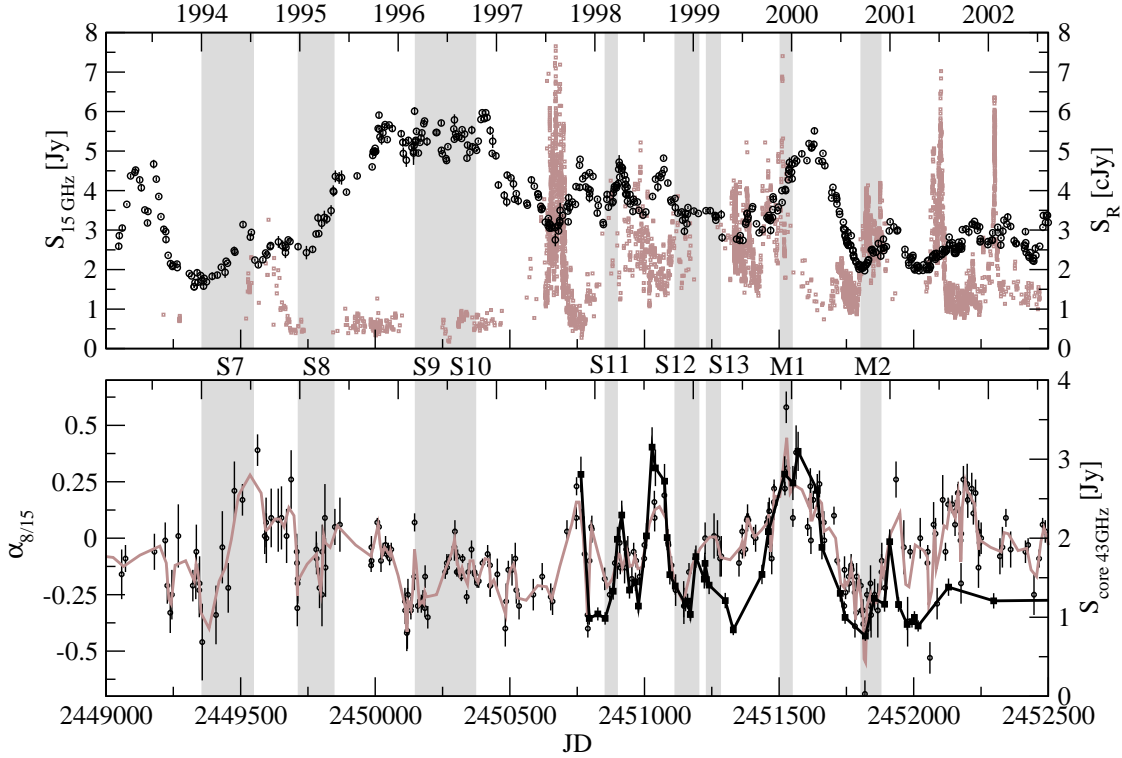}
\caption{Top: Optical (grey squares) and radio (black circles) light curves in
comparison with the ejection dates of new superluminal jet components, which are
highlighted as grey vertical bars. The width of each bar corresponds to the
uncertainty of the ejection date. Bottom: Ejection dates vs. radio spectral
index (black circles and grey trend-line) and the 43\,GHz core flux density
(black squares and line). The ejection dates are taken from
\cite{2000ApJS..129...61D}, \cite{2003MNRAS.341..405S} (S7 to S13), and
\cite{2004AIPC..714..167M} (M1 and M2). Only some of the new components seem to
be connected with events in the light curves.}\label{fig:spixzoom}
\end{figure*}

Since outbursts (especially at mm wavelengths) are often related to the ejection
of superluminal jet components (e.g,
\citealt{1998A&A...334..489O,2002A&A...394..851S}), we have compared our radio
and optical light curves and the spectral index, $\alpha_{8/15}$, with the
ejection dates of jet components reported in the literature.
Figure~\ref{fig:spixzoom} illustrates this comparison. There is no obvious
connection among the events, but some new components are led by an optical flare
or followed by an increase in the radio. We also tested whether we could find
any evidence for a connection between the ejections and flares by calculating
the variations in the flux density at different frequencies and the spectral
indices for times shortly before and after the ejection, but no clear trend could
be found. There is always a significant fraction of the components that does not
show the expected behaviour. However, this kind of analysis might be affected by
the relatively large errors in the dates of the earlier ejections (S7 to S10)
and the fact that some components appear very close together (S9/S10 and
S12/S13). Furthermore, the light curves are generally composites of multiple
events, some of which are declining in flux while others are rising. This can
render individual flares unnoticed in the total-flux light curves.

The best agreement can be seen for the two newest components M1 and M2. They
appear accompanied by an optical flare and are followed by a flare and
flattening of the spectrum at radio frequencies. Since previous works have also
reported a good agreement between the appearance of flares and the ejection of
components from S1 up to S7 (e.g.,
\citealt{1990ApJ...352...81M,1999ASPC..159...31A}), it might be that the dates
of S8 to S13 are not so well determined. Although S7, S11, and S12 can be
identified with minor radio flares and a flattening of the spectrum, there are
several large flares in the optical and radio light curves at 1997.6, 1998.4 and
1999.5 that seem to be unrelated to any new component. It will be interesting to
see if the ongoing densely sampled VLBI monitoring will find corresponding
components to the pronounced optical flares around 2001.5 and 2002.1. 

\begin{figure}[htbp]
\centering
\includegraphics[bb= 50 50 554 775,angle=-90,width=9cm,clip]{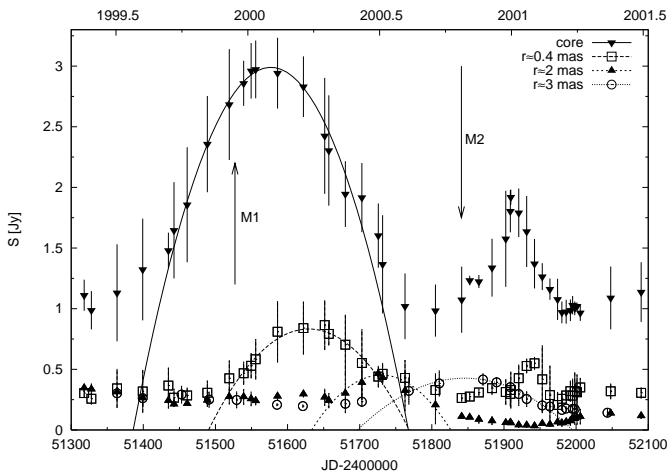}
\caption{VLBI light curves of the radio flare in early 2000. To better follow
the flare evolution in both time and space, all fluxes are converted to
22\,GHz according to their spectral index and combined in a three point average.
One can follow the propagation of the core outburst along the jet. The
ejection dates of new VLBI components are marked by arrows and correspond to M1 in
1999.95 and M2 in 2000.81 reported by
\cite{2003heba.conf..173M}.}\label{fig:2000flare}  
\end{figure}

As mentioned earlier, one should also expect to see a sign of the travelling jet
components in the different parts of the VLBI light curves. After a flare in the
VLBI core, a new component should brighten the jet after some time at every
measured distance while it passes by. Unfortunately, the jet flux density is
either too low or the time sampling is insufficient to reliably detect such
events. In addition, the ejection of several components or the appearance of
many flares in a short time can confuse the picture. A good example of the
evolution of a new jet component is M1 (Fig.~\ref{fig:spixzoom}) ejected in
1999.95 (JD=2451527). The corresponding flare is a nicely separated event and
from Fig.~\ref{fig:VLBAvstime} one can see that, some time after the peak in the
core, the inner jet ($r\approx 0.4$\,mas) brightens at 43\,GHz and 22\,GHz,
while later it appears at 2\,mas and finally at 3\,mas separation from the core.
Since not all frequencies were observed simultaneously and in order to see
better the evolution, we have converted the flux densities at 15\,GHz and
43\,GHz of the core and jet to 22\,GHz according to their measured spectral
index and show these together in Fig.~\ref{fig:2000flare}. The plotted data
points are three-point averages to smooth the residual differences, and the
error bars correspond to their standard deviations. Parabolic fits to the data
illustrate the appearance of the component at the different jet positions. The
calculated speed of the component from the peaks in the light curves is about 
(2--4)\,mas/yr or $\beta_{\rm app}\approx$~(12--19)\,$c$. This is somewhat
higher than the speeds of 5\,$c$ to $9\,c$ from kinematic studies reported thus
far
(\citealt{2000ApJS..129...61D,2003MNRAS.341..405S,2004ApJ...609..539K,2005AJ....130.1418J}),
but since we have combined all frequencies in order to be able to fit the peak,
our values might be more uncertain.

Another jet component ejection (M2) is reported for 2000.81 (JD=2451841;
\citealt{2003heba.conf..173M}) and the related radio flare is also visible in
Fig.~\ref{fig:2000flare}. There is a sharp core peak at JD=2451920 and some time
later a peak in the $r\approx0.4$\,mas light curve. Unfortunately, the dense
VLBI monitoring ends in early 2001 and one cannot follow the flare further.
However, both ejections are accompanied or led by a bright optical state, which
suggests that the correlated optical-radio variability observed here is also
related to the ejection of jet components.

During 1999 and 2000 BL\,Lac was the target of several {\it RXTE}
(\citealt{2004AIPC..714..167M}) and {\it BeppoSAX}
(\citealt{2002A&A...383..763R,2003A&A...408..479R,2003ApJ...596..847B})
observations. Interestingly, the source was caught in very different states. The
{\it RXTE} light curve covers nearly the whole period of 1999 to 2001 with dense
sampling. In particular, during the ejection in 2000.81 strong variability is
observed together with a softening of the X-ray spectrum
(\citealt{2004AIPC..714..167M}). This is also confirmed by the {\it BeppoSAX}
observations from October 31 to November 2, 2000 (JD$\sim$2451850;
\citealt{2003A&A...408..479R,2003ApJ...596..847B}), which show an unusually
strong soft spectral component between 0.2\,keV and about 10\,keV.
Unfortunately, the flare/ejection in 1999.95 was not covered by {\it RXTE},
while the {\it BeppoSAX} observations  of December 5--6 (JD$\sim$2451518)
obtained several days before the extrapolated ejection date showed no sign of a
soft component, but a hard spectrum similar to that observed in July 2000.
Another {\it BeppoSAX} observation on June 5--7, 1999 (JD$\sim$2451336) shows
again a small soft component between 0.2\,keV and about 3\,keV
(\citealt{2002A&A...383..763R}), which seems also visible as a small enhancement
in the {\it RXTE} light curve and spectrum (\citealt{2004AIPC..714..167M}). This
small X-ray flare is also followed by optical and radio flares, but no new jet
component has been reported. However, this might be due to confusion with the
two earlier ejections of components S12 and S13, which were separated by only
0.3\,yr.

Although further coordinated observations are needed to complete the picture, the
very well sampled period from 1999 to 2001 already reveals some very interesting
similarities in the variability behaviour from X-ray to radio wavelengths, which
are also visible as changes in the source VLBI structure. The missing
soft-spectrum component in the {\it BeppoSAX} observations of 1999.92 might
blear the claim of a correlation, but given that
the typical uncertainty in the ejection dates is about 0.2\,yr and the soft
X-ray tail is potentially a short-lived phenomenon, more observations yielding
better statistics are necessary.

\subsection{A possible jet scenario}\label{sec:jmod}

Due to the combination of the structural analysis using VLBI light curves and
the single-dish observations, we are confident that the emission from the VLBI
core, presumably close to the base of the jet, is responsible for the correlated
variability observed at optical and radio wavelengths, and possibly also up to
X-ray frequencies (\citealt{2004AIPC..714..167M}). The mechanisms that can
explain blazar variability are still the subject of discussion. Possibilities
suggested in the literature include shocks in jets
(\citealt{1985ApJ...298..114M,1985ApJ...298..296A,1996etrg.conf...45M}), and flares
introduced by the lighthouse effect from changes in the direction of forward
beaming caused, e.g., by helical trajectories  of plasma elements
(\citealt{1992A&A...255...59C}), by a precessing binary black-hole system
(\citealt{1980Natur.287..307B,1988ApJ...325..628S}), or by the rotation of a
helical jet
(\citealt{1998MNRAS.293L..13V,1999A&A...347...30V,2004A&A...419..913O}). Furthermore,
variability caused by colliding relativistic plasma shells has been considered
(\citealt{2001MNRAS.325.1559S,2004A&A...421..877G}).

Regardless of the variability origin, the VLBI observations of BL\,Lac indicate
a bent path of the jet. The trajectories of several jet components have been
well modelled and predicted by helical jet models
(\citealt{2000ApJS..129...61D,2003MNRAS.341..405S}). A bent jet path could also
explain the region of low emission around 1\,mas from the core noticed in
Sect.~\ref{sec:VLBIdata} (Fig.~\ref{fig:slices}). One can see strong variability
in both the core and the jet at core separations larger than 1\,mas, but never
between 0.7\,mas and 1\,mas. Since the spectral index does not show any evidence
for absorption, a natural explanation would be reduced Doppler boosting either
from a misalignment between the jet direction and our line of sight or from a
change in the jet speed.

A possible scenario that could account for the observed variability behaviour,
especially the changing variability pattern between optical and radio emission
on timescales of several years, would be a helical jet structure that is
precessing or rotating. Since the optical emission is most likely produced on
small scales, a small misalignment could beam the emission less favourably,
which could explain the low optical state with suppressed variability from 1981
to early 1997 (see Sect.~\ref{sec:timetest}). On the contrary, the radio
emission, coming from all along the jet, would be always strongly beamed in some
part of the helical path and therefore would not suffer the less intense beaming
of the optical component (e.g.,
\citealt{2004ApJ...615L...5R,2005ChJAA...5S.305R} and references therein). The
only beaming-suppressed region at radio frequencies would be the emission gap in
the inner jet. In this case the gap might move or disappear with time and could
be used by future VLBI studies to prove or disprove the rotating jet scenario.

\section{Summary}\label{conclusions}

We have presented an analysis of a VLBI data set that includes 108 images at
15\,GHz, 22\,GHz, and 43\,GHz obtained between 1995 and 2003. The aim has been
to disentangle the different components contributing to the single-dish radio
light curves. Inspection of the VLBI light curves corresponding to different
regions of the jet reveals that the radio single-dish light curves indeed
display the sum of the emission from the core and from some prominent jet
features. In some cases these jet features become as bright as, or even brighter
than, the core itself. This leads to a complication of the cross-correlation
between the radio light curves and other wavebands where the emission might come
from the core region alone.

We find that the spectral variability that is present in the single-dish light
curves can trace the variability of the VLBI core, and therefore enables a study
of the long-term variability of the radio core even when VLBI data are not
available. Using this result, we find a fair correlation between the variations
of the radio spectral index and the $R$-band optical light curve with radio
delays of about 50 to 180 days depending on the frequency separation. (Here we
attribute the delay measured from the spectral index between two frequencies to
the higher frequency. This seems justified as the flattening and also the peak
of the spectral index curve is mainly due to the earlier and faster rise of the
flare at the higher frequency.) Owing to the shorter time range and the sparser
sampling, the radio-optical correlation is weaker when considering the VLBI core
light curves at 22\,GHz and 43\,GHz directly, but the delays found are
comparable to those of the spectral indices. The resulting time lags are also
consistent with those from the cross-correlations performed by
\cite{2004A&A...424..497V}.

Assuming that the radio time lags are due to an optical depth effect of
synchrotron self-absorption and that flares propagate along the jet at a
constant speed, we obtain a power-law dependence of the time lag on frequency
that suggests that the jet does not maintain equipartition of magnetic and
particle energy densities as it expands. The radio-optical correlation is most
prominent before 1981 and after 1997. During the 16 years in between, we observe
only moderate variability in the optical and the correlation with the radio
almost disappears. Stirling et al.\ (2001) have suggested a precessing jet
nozzle for BL Lac to explain their observations, which may imply that the
optical variability is also orientation dependent. This in turn, depending on
the geometry, could also affect the correlation with the radio regime. This
would lead to a natural explanation of the changing appearance of correlated
variability: since the optical emission likely originates on small scales, a
minor misalignment could beam the emission less favourable. On the other hand,
the beamed radio emission could still repeatedly approach our line of sight
along its helical path farther down the jet, so the precession or rotation might
modify the amplitude of the variability pattern without leading to a vanishing
flux as in the optical bands (e.g., Rieger 2004, 2005). A bent jet path could
also explain the low emission region in the jet around 1\,mas from the core. A
movement or the disappearance of the gap in future VLBI studies, could
therewith be used to prove or disprove a rotation of the bent jet path.

\begin{acknowledgements}
We thank the referee, Steve Bloom, for his suggestions on improving the
paper. We are grateful to the group of the VLBA 2\,cm Survey and the group of the MOJAVE
project for providing their data. The VLBI observations were obtained using the
VLBA, which is an instrument of the National Radio Astronomy Observatory, a
facility of the National Science Foundation, operated under cooperative
agreement by Associated Universities, Inc.\ and is also based on observations
with the 100\,m radio telescope of the MPIfR (Max-Planck-Institut f\"ur
Radioastronomie) at Effelsberg. This research has made use of data from the
University of Michigan Radio Astronomy Observatory which has been supported by
the University of Michigan and the National Science Foundation. This work is
supported by the European Community's Human Potential Programme under contract
HPRCN-CT-2002-00321 (ENIGMA Network). The research of the Boston University
coauthors is supported in part by US National Science Foundation grant
AST-0406865.

\end{acknowledgements}

\bibliographystyle{aa} 
\bibliography{references}

\Online
\appendix

\setcounter{table}{0}
\begin{longtable}{lcrrcccc}
 \caption{Observing log. Given are the observing epoch, frequency, beam size and
orientation, total flux in the image (S$_{\rm tot}$), peak flux density (S$_{\rm
max}$), rms noise level (S$_{\rm rms}$), and the references where the data was
first published.}\\
 \label{tab:obslog}\\
   \hline
   \multicolumn{1}{c}{Epoch} &
   \multicolumn{1}{c}{$\nu$} &
   \multicolumn{2}{c}{Beam}  &
   \multicolumn{1}{c}{S$_{\rm tot} $} &
   \multicolumn{1}{c}{S$_{\rm max} $} &
   \multicolumn{1}{c}{S$_{\rm rms} $} &
   \multicolumn{1}{c}{Ref.} \\
   \multicolumn{1}{c}{} &
   \multicolumn{1}{c}{[GHz]} &
   \multicolumn{1}{c}{[mas x mas]} &
   \multicolumn{1}{r}{[$^\circ$]} &
   \multicolumn{1}{c}{[Jy]} &
   \multicolumn{1}{c}{[Jy]} &
   \multicolumn{1}{c}{[mJy]} &
   \multicolumn{1}{c}{} \\ \hline 
\endfirsthead

\multicolumn{7}{c}{{ \tablename\ \thetable{} -- continued from previous
column}}\\
\hline
   \multicolumn{1}{c}{Epoch} &
   \multicolumn{1}{c}{$\nu$} &
   \multicolumn{2}{c}{Beam} &
   \multicolumn{1}{c}{S$_{\rm tot} $} &
   \multicolumn{1}{c}{S$_{\rm max} $} &
   \multicolumn{1}{c}{S$_{\rm rms} $} &
   \multicolumn{1}{c}{Ref.} \\
   \multicolumn{1}{c}{} &
   \multicolumn{1}{c}{[GHz]} &
   \multicolumn{1}{c}{[mas x mas]} &
   \multicolumn{1}{r}{[$^\circ$]} &
   \multicolumn{1}{c}{[Jy]} &
   \multicolumn{1}{c}{[Jy]} &
   \multicolumn{1}{c}{[mJy]} &
   \multicolumn{1}{c}{} \\ \hline 
\endhead

\hline
\multicolumn{7}{c}{{Continued on next page}} \\
\endfoot

\hline
\endlastfoot
1995.27 & 15 & $0.50\times0.88$ & $2.9  $ & 3.30 & 1.79 & 2.4 & (1) \\
1995.96 & 15 & $0.53\times0.97$ & $0.9  $ & 5.01 & 2.04 & 2.5 & (1) \\
1996.38 & 15 & $0.51\times0.90$ & $-4.1 $ & 5.67 & 2.68 & 2.4 & (1) \\
1996.82 & 15 & $0.46\times0.86$ & $-2.6 $ & 3.84 & 1.77 & 3.2 & (1) \\
1997.19 & 15 & $0.58\times1.00$ & $7.1  $ & 3.21 & 1.49 & 1.5 & (1) \\
1997.66 & 15 & $0.56\times0.91$ & $12.4 $ & 2.97 & 1.70 & 1.5 & (1) \\
1998.18 & 15 & $0.50\times1.00$ & $8.2  $ & 2.68 & 1.06 & 3.8 & (1) \\
1998.73 & 15 & $0.57\times0.87$ & $-16.0$ & 4.07 & 2.48 & 0.7 & (5)     \\
1998.97 & 15 & $0.59\times0.71$ & $-3.0 $ & 2.98 & 1.54 & 0.7 & (5)     \\
1999.16 & 15 & $0.63\times0.78$ & $6.5  $ & 3.26 & 1.88 & 0.6 & (5)     \\
1999.40 & 15 & $0.60\times0.81$ & $6.2  $ & 2.38 & 1.30 & 0.7 & (5)     \\
1999.70 & 15 & $0.55\times0.77$ & $-14.1$ & 2.75 & 1.71 & 0.6 & (5)     \\
1999.85 & 15 & $0.59\times1.08$ & $-3.3 $ & 3.31 & 2.34 & 1.3 & (1) \\
2000.01 & 15 & $0.56\times0.79$ & $-16.1$ & 4.23 & 3.32 & 0.7 & (5)     \\
2000.31 & 15 & $0.55\times0.74$ & $-4.7 $ & 4.18 & 3.16 & 0.6 & (5)     \\
2000.49 & 15 & $0.56\times1.00$ & $16.7 $ & 3.23 & 1.96 & 0.5 & (4)      \\
2000.84 & 15 & $0.57\times0.97$ & $27.1 $ & 2.14 & 1.32 & 0.4 & (4)      \\
2000.99 & 15 & $0.54\times0.94$ & $-16.3$ & 2.59 & 1.82 & 1.1 & (1) \\
2001.00 & 15 & $0.47\times1.46$ & $-6.0 $ & 2.83 & 2.08 & 0.8 & (4)      \\
2001.09 & 15 & $0.49\times1.27$ & $-10.9$ & 2.46 & 1.79 & 1.1 & (4)      \\
2001.17 & 15 & $0.52\times1.37$ & $-1.8 $ & 1.92 & 1.35 & 0.9 & (4)      \\
2001.22 & 15 & $0.57\times1.02$ & $33.8 $ & 1.78 & 1.14 & 0.3 & (4)      \\
2001.25 & 15 & $0.54\times1.44$ & $-1.1 $ & 1.96 & 1.26 & 0.6 & (4)      \\
2001.97 & 15 & $0.48\times0.92$ & $1.6  $ & 2.34 & 1.44 & 0.7 & (1) \\
2002.03 & 15 & $0.32\times0.64$ & $-33.2$ & 2.62 & 1.40 & 0.4 & (6)       \\
2002.05 & 15 & $0.85\times1.15$ & $29.3 $ & 2.75 & 1.79 & 0.6 & (5)     \\
2002.45 & 15 & $0.52\times0.79$ & $-18.1$ & 2.18 & 1.24 & 0.7 & (2)     \\
2002.51 & 15 & $0.24\times0.55$ & $-20.4$ & 2.52 & 1.38 & 0.5 & (6)       \\
2003.10 & 15 & $0.51\times0.76$ & $-0.1 $ & 1.75 & 0.98 & 0.5 & (2)     \\
2003.24 & 15 & $0.29\times0.46$ & $-35.8$ & 1.73 & 0.80 & 0.3 & (6)       \\
2003.65 & 15 & $0.48\times0.71$ & $0.8  $ & 3.11 & 2.40 & 0.7 & (2)     \\
2004.44 & 15 & $0.53\times0.73$ & $-9.3 $ & 2.73 & 2.01 & 0.7 & (2)     \\
1997.94 & 22 & $0.39\times0.84$ & $-14.1$ & 2.79 & 1.54 & 1.2 & (4)      \\
1998.03 & 22 & $0.45\times0.87$ & $-8.0 $ & 3.06 & 1.51 & 1.1 & (4)      \\
1998.10 & 22 & $0.34\times0.85$ & $-11.3$ & 2.81 & 1.21 & 1.3 & (4)      \\
1998.18 & 22 & $0.36\times1.00$ & $1.9  $ & 2.90 & 1.42 & 1.6 & (4)      \\
1998.27 & 22 & $0.35\times0.85$ & $-8.2 $ & 6.60 & 3.36 & 2.2 & (4)      \\
1998.35 & 22 & $0.36\times0.83$ & $-7.4 $ & 3.17 & 1.58 & 1.5 & (4)      \\
1998.44 & 22 & $0.35\times0.87$ & $-8.6 $ & 2.79 & 1.48 & 1.4 & (4)      \\
1998.52 & 22 & $0.33\times0.97$ & $-13.6$ & 3.22 & 1.82 & 1.5 & (4)      \\
1998.61 & 22 & $0.36\times0.87$ & $-10.4$ & 3.97 & 2.71 & 2.1 & (4)      \\
1998.73 & 22 & $0.42\times0.64$ & $-17.8$ & 3.25 & 2.13 & 1.0 & (5)     \\
1998.82 & 22 & $0.35\times0.93$ & $-21.4$ & 3.30 & 1.88 & 1.4 & (4)      \\
1998.92 & 22 & $0.36\times0.88$ & $-14.8$ & 2.72 & 1.44 & 1.5 & (4)      \\
1998.97 & 22 & $0.48\times0.57$ & $-5.9 $ & 2.52 & 1.38 & 0.6 & (5)     \\
1999.03 & 22 & $0.38\times1.02$ & $-13.2$ & 3.72 & 2.07 & 1.1 & (4)      \\
1999.11 & 22 & $0.36\times0.82$ & $-8.3 $ & 2.82 & 1.79 & 1.6 & (4)      \\
1999.16 & 22 & $0.44\times0.56$ & $1.6  $ & 2.83 & 1.72 & 0.5 & (5)     \\
1999.41 & 22 & $0.41\times0.58$ & $-5.5 $ & 1.93 & 1.06 & 0.6 & (5)     \\
1999.70 & 22 & $0.39\times0.64$ & $-26.4$ & 2.07 & 1.37 & 0.7 & (5)     \\
2000.01 & 22 & $0.43\times0.59$ & $-15.7$ & 4.32 & 3.43 & 0.5 & (5)     \\
2000.31 & 22 & $0.43\times0.56$ & $-8.1 $ & 3.76 & 2.76 & 0.5 & (5)     \\
2000.49 & 22 & $0.40\times0.74$ & $12.8 $ & 3.20 & 2.02 & 1.3 & (4)      \\
2000.84 & 22 & $0.45\times0.69$ & $31.4 $ & 2.20 & 1.43 & 0.8 & (4)      \\
2001.00 & 22 & $0.37\times0.96$ & $-4.0 $ & 2.65 & 2.07 & 1.3 & (4)      \\
2001.09 & 22 & $0.38\times0.84$ & $-13.3$ & 2.38 & 1.65 & 1.5 & (4)      \\
2001.17 & 22 & $0.40\times0.94$ & $-1.9 $ & 1.69 & 1.20 & 1.1 & (4)      \\
2001.22 & 22 & $0.46\times0.78$ & $33.7 $ & 1.98 & 1.34 & 0.7 & (4)      \\
2001.25 & 22 & $0.39\times0.98$ & $-4.8 $ & 1.86 & 1.21 & 0.8 & (4)      \\
2001.59 & 22 & $0.37\times0.63$ & $-11.3$ & 1.88 & 1.34 & 0.7 & (5)     \\
2002.05 & 22 & $0.65\times0.92$ & $24.0 $ & 2.28 & 1.61 & 0.9 & (5)     \\
1997.86 & 43 & $0.17\times0.43$ & $-10.5$ & 3.67 & 2.74 & 2.1 & (4)      \\
1997.94 & 43 & $0.18\times0.44$ & $-10.2$ & 1.75 & 0.98 & 1.8 & (4)      \\
1998.03 & 43 & $0.23\times0.46$ & $-9.1 $ & 2.02 & 1.04 & 1.1 & (4)      \\
1998.10 & 43 & $0.17\times0.43$ & $-12.2$ & 1.93 & 0.97 & 1.6 & (4)      \\
1998.18 & 43 & $0.17\times0.53$ & $-2.7 $ & 2.30 & 1.33 & 1.8 & (4)      \\
1998.23 & 43 & $0.17\times0.33$ & $9.9  $ & 3.49 & 1.96 & 1.0 & (3)   \\
1998.27 & 43 & $0.18\times0.44$ & $-10.2$ & 3.56 & 2.17 & 1.9 & (4)      \\
1998.35 & 43 & $0.16\times0.44$ & $-10.5$ & 2.35 & 1.30 & 2.4 & (4)      \\
1998.41 & 43 & $0.16\times0.32$ & $-14.6$ & 2.70 & 1.35 & 0.6 & (3)   \\
1998.44 & 43 & $0.18\times0.44$ & $-9.4 $ & 2.01 & 1.13 & 2.1 & (4)      \\
1998.52 & 43 & $0.17\times0.57$ & $-15.6$ & 3.21 & 1.96 & 1.8 & (4)      \\
1998.58 & 43 & $0.16\times0.28$ & $1.1  $ & 4.28 & 2.98 & 0.8 & (3)   \\
1998.61 & 43 & $0.19\times0.43$ & $-10.6$ & 3.85 & 2.80 & 1.8 & (4)      \\
1998.70 & 43 & $0.17\times0.56$ & $-17.3$ & 3.88 & 2.65 & 1.4 & (4)      \\
1998.73 & 43 & $0.20\times0.32$ & $-25.9$ & 3.07 & 1.90 & 0.9 & (5)     \\
1998.76 & 43 & $0.19\times0.30$ & $-11.5$ & 2.50 & 1.44 & 0.7 & (3)   \\
1998.82 & 43 & $0.17\times0.47$ & $-19.3$ & 2.33 & 1.34 & 1.7 & (4)      \\
1998.92 & 43 & $0.21\times0.49$ & $-23.6$ & 1.99 & 1.18 & 1.4 & (4)      \\
1998.94 & 43 & $0.19\times0.30$ & $-13.4$ & 2.00 & 1.14 & 0.7 & (3)   \\
1998.97 & 43 & $0.21\times0.26$ & $-11.3$ & 1.86 & 0.99 & 0.7 & (5)     \\
1999.03 & 43 & $0.17\times0.60$ & $-11.8$ & 2.96 & 1.71 & 1.6 & (4)      \\
1999.11 & 43 & $0.17\times0.42$ & $-9.7 $ & 2.05 & 1.47 & 1.7 & (4)      \\
1999.12 & 43 & $0.19\times0.32$ & $-13.9$ & 2.71 & 1.64 & 0.6 & (3)   \\
1999.16 & 43 & $0.21\times0.27$ & $-1.1 $ & 2.31 & 1.38 & 0.6 & (5)     \\
1999.32 & 43 & $0.18\times0.26$ & $7.8  $ & 2.10 & 1.19 & 1.2 & (3)   \\
1999.41 & 43 & $0.23\times0.39$ & $-18.7$ & 1.50 & 0.84 & 0.7 & (5)     \\
1999.70 & 43 & $0.19\times0.35$ & $-30.0$ & 2.12 & 1.50 & 0.7 & (5)     \\
1999.76 & 43 & $0.20\times0.38$ & $-16.8$ & 2.73 & 2.00 & 1.2 & (3)   \\
1999.93 & 43 & $0.20\times0.38$ & $-15.9$ & 3.68 & 2.76 & 1.0 & (3)   \\
2000.01 & 43 & $0.23\times0.31$ & $-9.7 $ & 3.34 & 2.58 & 0.9 & (5)     \\
2000.06 & 43 & $0.19\times0.26$ & $7.2  $ & 4.21 & 3.00 & 1.7 & (3)   \\
2000.26 & 43 & $0.18\times0.28$ & $1.5  $ & 4.17 & 2.45 & 1.0 & (3)   \\
2000.31 & 43 & $0.19\times0.26$ & $2.0  $ & 2.95 & 1.80 & 0.9 & (5)     \\
2000.49 & 43 & $0.21\times0.37$ & $20.7 $ & 2.39 & 1.37 & 1.6 & (4)     \\
2000.54 & 43 & $0.18\times0.30$ & $-6.3 $ & 2.03 & 0.97 & 0.8 & (3)   \\
2000.75 & 43 & $0.19\times0.32$ & $-8.8 $ & 1.47 & 0.74 & 0.8 & (3)   \\
2000.84 & 43 & $0.23\times0.35$ & $32.5 $ & 1.88 & 1.23 & 1.7 & (4)     \\
2000.94 & 43 & $0.23\times0.33$ & $28.3 $ & 1.73 & 1.19 & 1.1 & (3)   \\
2001.00 & 43 & $0.18\times0.49$ & $-4.6 $ & 2.83 & 1.98 & 1.2 & (4)     \\
2001.09 & 43 & $0.18\times0.44$ & $-12.6$ & 1.94 & 1.18 & 1.7 & (4)     \\
2001.17 & 43 & $0.17\times0.49$ & $-4.4 $ & 1.46 & 0.92 & 1.5 & (4)     \\
2001.22 & 43 & $0.21\times0.36$ & $33.4 $ & 1.61 & 0.99 & 1.2 & (4)     \\
2001.25 & 43 & $0.19\times0.48$ & $-3.7 $ & 1.63 & 1.02 & 1.5 & (4)     \\
2001.28 & 43 & $0.17\times0.29$ & $-2.6 $ & 1.44 & 0.83 & 0.6 & (3)   \\
2001.59 & 43 & $0.26\times0.46$ & $13.1 $ & 1.99 & 1.48 & 1.0 & (5)     \\
2002.05 & 43 & $0.33\times0.57$ & $33.3 $ & 1.73 & 1.27 & 1.4 & (5)     \\
2003.82 & 43 & $0.20\times0.38$ & $-22.1$ & 1.88 & 1.27 & 0.7 & (5)     \\ 
\end{longtable}
Note:~(1) VLBA 2\,cm Survey
(\citealt{1998AJ....115.1295K,2002AJ....124..662Z,2004ApJ...609..539K}); (2) 
MOJAVE (\citealt{2005AJ....130.1389L}); (3)
\cite{2001ApJ...556..738J,2005AJ....130.1418J}, \cite{2003MNRAS.341..405S}; (4)
G\'omez et al.\ \& Agudo et al.\ priv.\ comm.; (5)
\cite{2000ApJS..129...61D}, \cite{2005ApJ...623...79M}; (6) this work.

\end{document}